\documentclass{article}

\usepackage{natbib}%
\setcitestyle{authoryear}
\usepackage[figuresright]{rotating}

\usepackage{url}
\usepackage{algorithmic}
\usepackage{algorithm}

\usepackage{amsmath,amsthm}

\theoremstyle{definition}

\theoremstyle{remark}

\usepackage{tikz}
\usetikzlibrary{automata}
\usepackage{mathtools}
\usepackage{amssymb}
\usepackage{makecell}
\usepackage{booktabs}
\usepackage{authblk}
\usepackage{arxiv}
\usepackage{enumitem}
\usepackage{caption}
\usepackage{subcaption}
\usepackage{arydshln}

\newcommand\blfootnote[1]{%
  \begingroup
  \renewcommand\thefootnote{}\footnote{#1}%
  \addtocounter{footnote}{-1}%
  \endgroup
}

\begin{document}

\blfootnote{This preprint has not undergone any post-submission improvements or corrections. The Version of Record of this article is 
published in the Journal of Combinatorial Optimization, and is available online at https://doi.org/10.1007/s10878-026-01410-x. SharedIt view-only version: https://rdcu.be/e7mHq}

\title{Online Dispatching and Routing for Automated Guided Vehicles in Pickup and Delivery Systems on Loop-Based Graphs}
\setshorttitle{Online Dispatching and Routing for AGVs}

\author[1,$\ast$]{Louis Stubbe}
\author[1]{Jens Goemaere}
\author[1]{Jan Goedgebeur}

\affil[ ]{louis.stubbe@kuleuven.be, jens.goemaere@gmail.com, jan.goedgebeur@kuleuven.be}
\affil[1]{Department of Computer Science KU Leuven campus Kulak, Etienne Sabbelaan 53, Kortrijk 8500, Belgium}
\affil[$\ast$]{Author to whom all correspondence should be addressed (e-mail: louis.stubbe@kuleuven.be).}

\maketitle

\begin{abstract}
Automated guided vehicles (AGVs) are widely used in various industries, and scheduling and routing them in a conflict-free manner is crucial to their efficient operation. We propose a loop-based algorithm that solves the online, conflict-free scheduling and routing problem for AGVs with any capacity and ordered jobs in loop-based graphs. The proposed algorithm is compared against an exact method, a greedy heuristic and a metaheuristic. We experimentally show, using theoretical and real instances on a model representing a real manufacturing plant, that this algorithm either outperforms the other algorithms or gets an equally good solution in less computing time.
\end{abstract}


\section{Introduction}\label{sec:Introduction}
Automated guided vehicles (AGVs) can be used to autonomously transport materials from point A to B without requiring any human intervention. This makes them an effective and flexible solution for industrial applications such as manufacturing, warehousing and logistics. However, scheduling and routing AGVs in a dynamic environment presents a challenging problem, particularly when multiple AGVs share the same space. Furthermore, real-world applications might require fast response times or necessitate the problem to be handled online. As described by \cite{le2006review}, all jobs that need to be scheduled are known in advance in an offline scenario. This is not the case in an online scenario, where jobs may be added over time.

More specifically, the schedule determines when and which AGVs should start or finish jobs based on some overarching problem. For example, a good schedule for the AGVs responsible for delivering pallets of new material to a machine, would ensure uninterrupted operation by always keeping it supplied. A routing solution then decides what route the AGVs should take to accomplish the tasks given by the schedule. As a result, a conflict-free scheduling and routing solution is one that satisfies the overarching problem  while adhering to various constraints such as avoiding AGV collisions and respecting location capacity limitations.

In this paper, we propose an online algorithm that integrates conflict-free scheduling and routing for AGVs in loop-based graphs. This algorithm is designed such that, to the best of our knowledge, it utilizes the underlying structure of loop-based graphs. AGVs may have any capacity and jobs may need to be completed in a particular order. We compare our algorithm to existing solving strategies \citep{MURAKAMI2020106270,ZhengYan2014Atsa} adapted for the specific problem, in a synthetic offline scenario and an online one based on historical data of a real manufacturing plant. Note that we did not replicate the exact implementations from the referenced papers. Rather, we adapted their general methodologies to the best of our abilities for the specific problems at hand, such as using a mixed-integer programming (MIP) formulation or tabu search algorithm. We demonstrate that our approach provides a better solution or a similar one in less computing time in most cases.

This paper is organized as followed. In Section~\ref{sec:Literature review}, existing research is discussed. Section~\ref{sec:Prolem statement} provides an informal and a formal description of the problem and introduces corresponding notation. Section~\ref{sec:Alg} presents an exact method, two heuristics and a metaheuristic that can solve the stated problem. Using artificial problem instances and a model representing a real plant as described in Section~\ref{sec:modeling}, computational experiments with these algorithms are reported in Section~\ref{sec:Results}. Finally, Section~\ref{sec:Threats to validity} lists potential threats to validity and how we mitigated them, while Section~\ref{sec:conclusion} presents some conclusions and potential future works.

\section{Literature review} \label{sec:Literature review}
Past research usually focuses on either the scheduling- or the routing problem but not both at the same time. For example, the work by~\cite{RashidiHassan2011Acaa} proposes an exact method based on minimum cost flow to solve the scheduling problem and an incomplete but fast alternative that can be used in online scenarios. A genetic algorithm to solve the offline scheduling problem was described by~\cite{JeraldJ.2006Ssop}.

Algorithms that solve the routing problem are also quite diverse. A first example is the work by ~\cite{FazlollahtabarHamed2010Mpat}, which solves it in an offline scenario using an exact method based on mathematical programming. Another example is the work by \cite{Tavakkoli-MoghaddamR.2008Pmit}, which proposes a metaheuristic that minimizes inter-loop and intra-loop flow simultaneously.

When solving both problems simultaneously, a decomposition approach is common. This usually involves solving the master scheduling problem with one method, constrained by the lower level routing problem with some other method. An example of this approach involves the work by \cite{NishiTatsushi2011Abda}, which uses a mixed integer formulation to find a lower bound for the master problem and a corresponding upper bound for the sub problem. Cuts are then used to exclude previously feasible solutions. A similar strategy was described by \cite{correa2007scheduling}, which uses constraint programming for the master problem and mixed integer programming for the sub problem. Both of these also assume the problem is offline. A parallel approach was proposed by \cite{AmirteimooriArash2023Aphf}, which creates multiple partial solutions and completes them in parallel.

Some examples where both problems are solved simultaneously without a decomposition strategy include the work by \cite{FazlollahtabarHamed2015Mmfd}, which constructs an initial feasible solution. This is then improved using dynamic programming. An approach using tabu search was implemented by~\cite{ZhengYan2014Atsa} and verified to be near-optimal using an improved lower bound calculation. Finally, the heuristic approach by~\cite{EbbenMark2004Moct} is able to construct an initial infeasible solution based on a relaxed problem, convert it to a feasible one and then further optimize it. All of these also assume the problem is offline.

The works by \cite{DrotosMarton2021Sacc} using two improvement strategies and \cite{MalopolskiWaldemar2018Asac} using a square layout with average speeds do provide a conflict-free scheduling and routing solution in an online setting under mild assumptions, but are non-capacitated and jobs must be completable in any order. Some capacitated solutions are the works by \cite{MiyamotoToshiyuki2016Lars} using a local and random search algorithm and the flow based solution of \cite{MURAKAMI2020106270}. However, these still assume jobs can be completed in any order.

Our proposed solution mainly differs from existing work by solving the both the scheduling and routing problem simultaneously in an online environment. AGVs may have any capacity and jobs may need to be completed in a particular order. As a drawback, we make stronger assumptions about the structure of the graph.

\section{Problem statement}\label{sec:Prolem statement}
\begin{figure}
	\centering
	\begin{tikzpicture}[darkstyle/.style={circle,draw,fill=gray!40,minimum size=10}]
		\pgfmathtruncatemacro{\xn}{4}
		\pgfmathtruncatemacro{\yn}{4}
		\pgfmathsetlengthmacro{\xspace}{1.7cm}
		\pgfmathsetlengthmacro{\yspace}{1.3cm}
		
		\pgfmathtruncatemacro{\xnm}{\xn-1}
		\pgfmathtruncatemacro{\ynm}{\yn-1}
		\pgfmathtruncatemacro{\ymid}{round(\yn/2)}
		\pgfmathtruncatemacro{\ymidp}{\ymid+1}
		\pgfmathtruncatemacro{\ymidm}{\ymid-1}
		
		\foreach \x in {0,...,\xnm}
		\foreach \y in {0,...,\yn} 
		\node [darkstyle]  (\x\y) at (\xspace*\x,\yspace*\y) {};
		
		\foreach \y in {0,...,\ymidm} 
		\node [darkstyle]  (\xn\y) at (\xspace*\xn,\yspace*\y) {};
		
		\node [darkstyle]  (\xn\ymid) at (\xspace*\xn,\yspace*\ymid) {S};
		
		\foreach \y in {\ymidp,...,\yn} 
		\node [darkstyle]  (\xn\y) at (\xspace*\xn,\yspace*\y) {};
		
		\foreach \y [count=\yi] in {0,...,\ynm}
		\path [->](\xn\yi) edge node[left] {} (\xn\y);
		
		\foreach \x in {0,...,\xnm}
		\foreach \y [count=\yi] in {0,...,\ynm}
		\path [->](\x\y) edge node[left] {} (\x\yi);
		
		\foreach \x [count=\xi] in {0,...,\xnm}
		\path [->](\xi0) edge node[left] {} (\x0);
		
		\foreach \x [count=\xi] in {0,...,\xnm}
		\path [->] (\x\yn) edge node[left] {} (\xi\yn);	
	\end{tikzpicture}
	\caption{A directed graph representing the general layout of a manufacturing plant that mainly consists of loops and that has a single stockroom marked S. All nodes have a directed edge connecting that node to itself, which is not drawn.}
	\label{fig:generalLayout}
\end{figure}
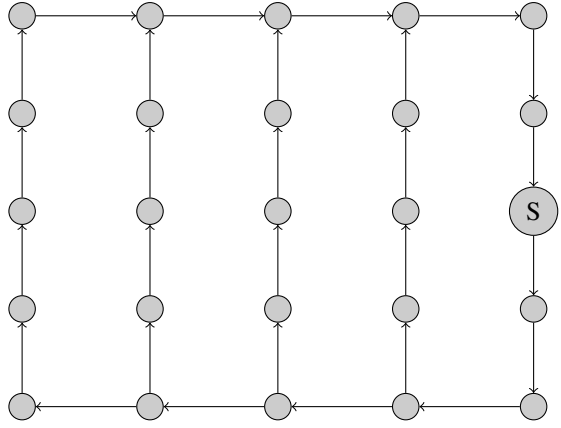

Consider a manufacturing plant with a single stockroom where pallets with new materials can be received or empty ones can be delivered. Several stations are placed throughout the plant where AGVs can deliver or pick up these pallets. We can represent this using a directed graph where the nodes correspond to the different delivery and pickup points. A \emph{cycle} is a path that starts and ends in the same node. A \emph{loop} will be defined as a cycle that starts or ends in the stockroom. We will assume the plant has a layout that does not allow overtaking and that any cycle must also be a loop. Note that this means deadlocks are effectively impossible. Standing still will be represented by the edge that connects a node to itself. An example of a directed graph that fulfills these constraints, is given in Fig. \ref{fig:generalLayout}. As a final simplification, we will assume time is discrete and that a move from one node to the next takes exactly one time step.

Operators working in the manufacturing plant can request the delivery of new pallets with materials, the removal of empty ones or a combination of these two in any node that is not the stockroom. Note that in this combined case, these two paired jobs must be completed in a particular order. More specifically, the new pallet cannot be delivered before the empty one has been removed. The requests are added throughout the day and are not known in advance. As a result, this problem is online.

The manufacturing plant is more efficient if operators have to wait as little as possible between requesting new materials and the pallets being delivered. This means a good solution tries to minimize the time between a request being made and it being completed, i.e. the \emph{job completion time}, specifically for the requests that bring new materials.

\subsection{Mathematical formulation - Offline}
\label{sec:MathForm}
This subsection provides a mathematical formulation for the offline problem statement. The online version where not all jobs are known and some of them may be partially completed is described in Subsection \ref{sec:mathOnline}.

\subsubsection{Problem parameters}
\label{subsubsec:param}
The following problem parameters and sets are used:

\begin{description}
	\item[Problem parameters]
\end{description}
\begin{description}[style=multiline]
	\item[$V$] the set of nodes
\item[$E$] the set of edges
\item[$A$] the set of AGVs
\item[$J$] the set of all jobs
\item[$J_n$] the set of jobs that bring new materials
\item[$J_b$] the set of jobs that cannot be unloaded before a job $j \in J$ has been loaded, i.e. the set of \emph{blocked} or incompletable jobs
\item[$O_{v}$] the set of outgoing edges $e_{\texttt{out}}$ for some node $v \in V$
\item[$I_{v}$] the set of incoming edges $e_{\texttt{in}}$ for some node $v \in V$
\end{description}

\begin{description}
	\item[Problem sets]
\end{description}
\begin{description}[style=multiline]
	\item[$H$] an integer indicating the time horizon
	\item[$C_{V, v}$] the capacity for some node $v \in V$, i.e. at most $C_{V, v}$ AGVs can be in node $v$ at any given time
	\item[$C_{E, e}$] the capacity for some edge $e \in E$, i.e. at most $C_{E, e}$ AGVs can use edge $e$ at any given time
	\item[$C_{A, a}$] the capacity for some AGV $a \in A$, i.e. AGV $a$ can only carry $C_{A, a}$ pallets at a time
	\item[$Q_a$] the initial node $v \in V$ for AGV $a \in A$
	\item[$S_j$] the start node $v \in V$ for job $j \in J$
	\item[$F_j$] the end node $v \in V$ for job $j \in J$, assuming all jobs either start or end in the stockroom $s \in V$: $\forall j \in J: \left( S_j = s \veebar F_j = s \right)$
	\item[$(v, w)$] the directed edge $(v, w) \in E$ connecting node $v \in V$ to node $w \in V$, assuming all nodes have an edge connecting that node to itself, which represents not moving: $\forall v \in V: \exists (v,v)$
	\item[$B(j)$] the job $B(j) \in J$ that must be loaded before the given job $j \in J_b$ can be unloaded
\end{description}

\subsubsection{Decision variables}
The following decision variables are used: 

\begin{description}
	\item[Decision variables]
\end{description}
\begin{description}[style=multiline]
	\item[$P_{t, a, e}$] a binary variable indicating the \emph{position}, i.e. whether or not AGV $a$ used the edge $e$ at time step $t$. Not moving is considered to be using the edge that connects a node to itself
	\item[$L_{t, a, j}$] a binary variable indicating \emph{loading}, i.e. whether or not the pallet required for job $j$ was loaded by AGV $a$ at time step $t$
	\item[$U_{t, a, j}$] a binary variable indicating \emph{unloading}, i.e. whether or not the pallet required for job $j$ was unloaded by AGV $a$ at time step $t$.
\end{description}

\subsubsection{Constraints}
AGVs must follow a continuous path throughout the graph. This means they can only use a single edge at any given time step \eqref{eq:mov1} and an outgoing edge for some node $v$ can only be used if the previous time step used an incoming edge for this $v$ \eqref{eq:mov2}. They must also adhere to the edge- \eqref{eq:gcap1} and node-capacity \eqref{eq:gcap2} of the graph. Finally, they must start in the correct initial node \eqref{eq:mov3}.
\begin{align}
		\forall a \in A, 0 \leq t \leq H: \quad&\sum_{\substack{e \in E}} P_{t,a,e} = 1 \label{eq:mov1} \\
		\forall a \in A, \forall v \in V, 0 \leq t \leq H-1: \quad&\sum_{\mathclap{\substack{e_{\texttt{in}} \in I_{v}}}} P_{t,a,e_{\texttt{in}}} = \sum_{\mathclap{\substack{e_{\texttt{out}} \in O_{v}}}} P_{t+1, a,e_{\texttt{out}}} \label{eq:mov2} \\
		\forall e \in E, 0 \leq t \leq H: \quad&\sum_{\substack{a \in A}} P_{t,a,e} \leq C_{E, e} \label{eq:gcap1} \\
		\forall v \in V, 0 \leq t \leq H: \quad&\sum_{\substack{a \in A}}\sum_{\substack{e_{\texttt{in}} \in I_v}} P_{t,a,e_{\texttt{in}}} \leq C_{V, v} \label{eq:gcap2} \\
		\forall a \in A: \quad&\sum_{\substack{e_{\texttt{out}} \in O_{Q_a}}} P_{0,a,e_{\texttt{out}}} = 1 \label{eq:mov3}
\end{align}

\noindent Jobs must be assigned to \eqref{eq:job1} and completed by \eqref{eq:job2} exactly one AGV at some point in time. A job assigned to some AGV must also be completed by this AGV by first assigning it and then completing it \eqref{eq:job3}.
\begin{align}
	\forall j \in J: \quad&\sum_{\substack{a \in A}} \sum_{t = 0}^H L_{t,a,j} = 1 \label{eq:job1}\\
	\forall j \in J: \quad&\sum_{\substack{a \in A}} \sum_{t = 0}^H U_{t,a,j} = 1 \label{eq:job2}\\
	\forall a \in A, \forall j \in J, 0 \leq t \leq H: \quad&\sum_{t' = 0}^t L_{t',a,j} \geq \sum_{t' = 0}^t U_{t',a,j} \label{eq:job3}
\end{align}

\noindent Assigning or completing a job requires that the chosen AGV remains in the designated start- or end node for one time step to load or unload the pallet (\ref{eq:job4}, \ref{eq:job5}). Since this loading and unloading takes a time step, assigning or completing more than one job at a given time is not allowed \eqref{eq:job7}. An AGV can only carry a pre-specified number of pallets at a given time \eqref{eq:job6}.
\begin{align}
		\forall a \in A, \forall j \in J, 0 \leq t \leq H: \quad&P_{t,a, E(S_{j}, S_{j})} \geq L_{t, a, j} \label{eq:job4} \\
		\forall a \in A, \forall j \in J, 0 \leq t \leq H: \quad&P_{t,a , E(F_j, F_j)} \geq U_{t, a, j} \label{eq:job5}\\
		\forall a \in A, 0 \leq t \leq H: \quad&\sum_{\substack{j \in J}} L_{t, a, j} + U_{t, a, j} \leq 1 \label{eq:job7} \\
		\forall a \in A, 0 \leq t \leq H: \quad&\sum_{t' = 0}^t \sum_{\substack{j \in J}} L_{t', a, j} - U_{t', a, j} \leq C_{A, a} \label{eq:job6}
\end{align}

\noindent If an old pallet needs to be removed and a new one needs to be delivered, then the delivery is not possible before the old one has been loaded \eqref{eq:job8}. There is only enough space in a node for one AGV to load or unload at a given time step (\ref{eq:job9}, \ref{eq:job10}).

\begin{align}
		\forall j \in J_b,\, 0 \leq t \leq H: \quad&\sum_{t' = 0}^t \sum_{\substack{a \in A}} U_{t',a,j} \leq \Biggr. \sum_{t' = 0}^t \sum_{\substack{a \in A}} L_{t',a,B(j)} \label{eq:job8} \\
		\forall j \in J,\, 0 \leq t \leq H: \quad&\sum_{\substack{a \in A}}\; \left( \sum_{\substack{j' \in J:\\ S_{j} = S_{j'}}} L_{t, a, j'} + \sum_{\substack{j' \in J:\\ S_{j} = F_{j'}}} U_{t, a, j'} \right) \leq 1 \label{eq:job9} \\
		\forall j \in J,\, 0 \leq t \leq H: \quad&\sum_{\substack{a \in A}}\; \left( \sum_{\substack{j' \in J:\\ F_j = S_{j'}}} U_{t, a, j'} + \sum_{\substack{j' \in J:\\ F_j = F_{j'}}} L_{t, a, j'} \right) \leq 1 \label{eq:job10}
\end{align}

\noindent The optimal solution for the problem has a minimal job completion time for jobs that bring new materials. The objective function is given in Equation \eqref{eq:goal}. 

\begin{equation}
	\min \left(\sum_{\substack{a \in A}} \sum_{\substack{j \in J_n}} \sum_{t = 0}^H t \cdot U_{t,a,j} \right) \label{eq:goal}
\end{equation}

\subsection{Mathematical Formulation - Online}
\label{sec:mathOnline}
Here, we provide a mathematical formulation of the online problem statement. One can easily see that the difference between the offline and online problem is the initial state. All the AGVs start at the 0-node in the offline case, whereas in the online case, certain AGVs can already be assigned to some jobs. More concretely, this will be done by extending the original formulation in Subsection \ref{sec:MathForm} with extra problem parameters and constraints and by relaxing some other constraints. 

Note that an optimal solution for a certain online time frame may not result in an optimal solution for the overarching problem. A specific choice or assignment may seem good now, but can hinder a better solution at a later time. For example: assume there are two AGVs that have a capacity of two slots and two combinable jobs. If we optimize Equation \eqref{eq:goal}, then we find the solution in which each AGV does one job. However, if we now add a third job at a later time, there is no longer an available AGV to complete it. This would not have happened if we used one AGV for both jobs at the start.

\subsubsection{Problem parameters}
The following extra parameters are required:
\begin{description}
	\item[Extra problem parameters]
	\item[$X$] the set of jobs $j \in J$ that were assigned but not fully completed
	\item[$A_j$] the AGV $A_j \in A$ that was assigned the job $j \in X$
\end{description}

\subsubsection{Constraints}
Some AGVs may already be carrying pallets from a previous time frame, which means they should be assigned immediately. Constraints regarding loading, which were Equations \eqref{eq:job5}, \eqref{eq:job7}, \eqref{eq:job9} and \eqref{eq:job10}, should exclude these pallets. This respectively results in the following constraints:
\begin{align}
		\forall j \in X: \quad&L_{0,A_j,j} = 1 \label{eq:online1}\\
		\forall a \in A, \forall j \in J \setminus X, 0 \leq t \leq H: \quad&P_{t, a, E(F_j, F_j)} \geq U_{t, a, j} \label{eq:seq2} \\
		\forall a \in A, 0 \leq t \leq H: \quad&\sum_{\substack{j \in J \setminus X}} L_{t, a, j} + \sum_{\substack{j \in J}} U_{t, a, j} \leq 1 \label{eq:seq3} \\
		\forall j \in J,\, 0 \leq t \leq H: \quad&\sum_{\substack{a \in A}} \left( \sum_{\substack{j' \in J \setminus X:\\ S_{j} = S_{j'}}} L_{t, a, j'} + \sum_{\substack{j' \in J:\\ S_{j} = F_{j'}}} U_{t, a, j'} \right) \leq 1 \label{eq:seq4} \\
		\forall j \in J,\, 0 \leq t \leq H: \quad&\sum_{\substack{a \in A}} \left( \sum_{\substack{j' \in J:\\ F_j = S_{j'}}} U_{t, a, j'} + \sum_{\substack{j' \in J \setminus X:\\ F_j = F_{j'}}} L_{t, a, j'} \right) \leq 1 \label{eq:seq5}
\end{align}

\section{Algorithms}
\label{sec:Alg}
This section describes four algorithms to solve the scheduling and routing problem: an exact method, a greedy- and loop-based heuristic and a metaheuristic. Since job requests can be added throughout time, they need to be able to deal with requests that were already in transit due the solution of a previous scheduling period. As a consequence, a single period should not take too much time, since more jobs may have been added during the calculation. 

\subsection{Exact method}
\label{sec:MIP}
In order to solve the model as described in Subsection~\ref{sec:mathOnline} to optimality using mixed integer programming, a value for the time horizon $H$ needs to be set. This was done using one of the heuristics in Subsection~\ref{sec:heuristic}, as their solutions give an upper bound on the time horizon for the optimal solution.

\begin{algorithm}[tb]
	\caption{Base heuristic}
	\label{alg:base}
	\textbf{Input}: The AGVs and jobs for this time window and the solution of the previous time window.\\
	\textbf{Output}: A solution for this time window of the scheduling and routing problem.\\
	\vspace{-1em}
	\begin{algorithmic}[1]
		\FOR{each started but unfinished job $j$ in previous time window}
		\STATE Reassign $j$ to the AGV it was already assigned to
		\ENDFOR
		\STATE Let time step $t=0$
		\WHILE{a job is not completed at $t$}
		\FOR{AGV $a \in A$}
		\IF {$a$ is idle at $t$}
		\STATE Try to assign jobs
		\ENDIF
		\ENDFOR
		\STATE $t \gets t + 1$
		\ENDWHILE
		\STATE \textbf{return} solution
	\end{algorithmic}
\end{algorithm}

\subsection{Heuristics}
\label{sec:heuristic}
Two heuristics were implemented to solve the optimization problem: A greedy heuristic that mirrors the currently used implementation by the real manufacturing plant and a loops heuristic that tries to exploit the shape of the graph. Both of them are based on a generic heuristic, described in Algorithm~\ref{alg:base} to decide when to add jobs to a given AGV. Information about the previous scheduling period is dealt with by letting an AGV finish their current loop and the jobs on this loop. The jobs that were not on this loop are then unassigned and added to the set of uncompleted jobs.

\subsubsection{Greedy heuristic}
\label{sec:Greedy}
The greedy heuristic works by assigning one job or job pair to the first available AGV and by using the shortest possible route using Dijkstra's algorithm to complete this job or job pair through the graph assuming it does not cause any conflicts. If this assignment does cause a conflict, then the AGV waits in its current node for one time step instead.

\begin{algorithm}[tb]
	\caption{Loops heuristic}
	\label{alg:loops}
	\textbf{Input}: A partial solution.\\
	\textbf{Output}: A partial solution with extra jobs assigned to the given AGV if possible.\\
	\vspace{-1em}
	\begin{algorithmic}[1]
		\STATE Calculate $L$: the loops in the graph
		\FOR{$j$ in $J$}
		\STATE Calculate $L_j$: the loops that contain $j$
		\ENDFOR
		\STATE Let $B$, the best job-loop combination, be \verb*|null|
		\FOR{$j$ in $J$}
		\STATE Let $S = L_{j}$, $i = 1$
		\STATE Let R be $J \setminus {j}$, \\sorted by job pairs and travel time
		\WHILE{$S \cap L_{R_i} \neq \emptyset$ and\\ no constraints violated}
		\STATE $S \gets S \cap L_{R_i}$
		\STATE $i \gets i + 1$
		\ENDWHILE
		\IF{Multiple $B$}
		\STATE $B \gets $ the shortest loop in $B$ 
		\ENDIF
		\IF{$S$ is better than $B$}
		\STATE $B \gets S$
		\ENDIF
		\ENDFOR
		\STATE Apply $B$ to the partial solution
		\STATE \textbf{return} the updated partial solution
	\end{algorithmic}
\end{algorithm}

\begin{algorithm}[tb]
	\caption{Calculate loops}
	\label{alg:floodfill}
	\textbf{Input}: A graph $G$ with a start node $s$.\\
	\textbf{Output}: A collection of loops $L$:.\\
	\vspace{-1em}
	\begin{algorithmic}[1]
		\STATE $X$ $\gets$ \{[$s$]\}
		\STATE $L$ $\gets$ \{\}
		\WHILE{$X$ not empty}
		\STATE Pick a partial loop $l \in X$
		\STATE $X \gets X \setminus l$
		\WHILE{$l[-1]$ has 1 outgoing connection to $c$ in $G$ and $l$ is not a full loop}
		\STATE $l \gets l + c$
		\ENDWHILE
		\FORALL{outgoing connections $c$ in $G$ from $l[-1]$ and $l$ is not a full loop}
		\STATE $X \gets X \cup (l + c)$				
		\ENDFOR
		\IF{$l$ is a full loop}
		\STATE $L \gets L \cup l$
		\ENDIF
		\ENDWHILE
	\end{algorithmic}
\end{algorithm}

\subsubsection{Loops heuristic}
The loops heuristic tries to exploit the shape of the graph in Fig.~\ref{fig:generalLayout}, which mainly consists of loops. As described in the MIP formulation in Subsection \ref{subsubsec:param}, a job starts at a node $S_j$ and ends at $F_j$. The set of loops $L_j$ corresponding to job $j$ contains all loops that include both $S_j$ and $F_j$. 

The heuristic tries to combine as many jobs as possible by keeping track of the set of loops that all currently chosen jobs are part of. For example, consider two jobs $j_1$ and $j_2$ with corresponding loop sets $L_{j_1}$ and $L_{j_2}$. If $L_{j_1} \cap L_{j_2}$ is not empty, then $j_1$ and $j_2$ share at least one loop and are combinable. This principle can also be applied to groups of jobs, which means it can be used to construct a group that shares at least one loop. In a second stage, the algorithm verifies whether assigning these jobs would exceed the AGV's capacity or if it violates a paired job constraint.

Algorithm~\ref{alg:loops} is a more precise description of how the loops heuristic tries to assign jobs. Note that this is effectively a depth first search where only a single branch is checked after the first level. The loops of the graph are calculated using a modified flood fill algorithm as described by Algorithm~\ref{alg:floodfill}. The order in which jobs are added is determined by whether or not the job is part of a pair and the minimal travel distance to complete the job. The following criteria are used to determine whether one solution is better than a different one:
\begin{itemize}
	\item $R_1$: The amount of assigned jobs.
	\item $R_2$: The amount of assigned jobs that must be loaded before some other job can be unloaded. 
	\item $R_3$: The length of the chosen path.
	\item $R_4$: The slot usage.
\end{itemize}

Solutions are compared based on $R_1$, and if tied, $R_2$ is considered, continuing in sequence until a distinction is made.

\begin{algorithm}[tb]
	\caption{Tabu search}
	\label{alg:tabu}
	\textbf{Input}: A feasible solution.\\
	\textbf{Output}: A feasible solution with less time steps.
	
	\begin{algorithmic}[1]
		\WHILE{none of the stop-criteria are fullfilled}
		\IF{current solution is feasible}
		\STATE Save current solution
		\STATE Remove one time step
		\ENDIF
		\STATE Let bestMove be \verb*|null| 
		\FOR{move in possible moves}
		\STATE Calculate cost
		\IF{(cost $<$ cost of bestMove and !tabu(move)) or aspiration criteria}
		\STATE bestMove $\gets$ move
		\ENDIF
		\ENDFOR
		\STATE Apply bestMove to current solution
		\STATE Make reverse(bestMove) tabu
		\ENDWHILE
		\STATE \textbf{return} the best feasible solution
	\end{algorithmic}
\end{algorithm}

\subsection{Tabu search}
\label{sec:TS}
The general concept behind \emph{tabu search (TS)} as described by \cite{glover1986future}, is to explore the solution space by moving between adjacent and possibly infeasible solutions, while keeping track of previously visited ones. A more detailed description can be found in Algorithm~\ref{alg:tabu}. 

A solution is represented as a route matrix with each row representing the path of an AGV at a given time step as a sequence of nodes $v$. This means entry $(i,j)$ would contain the location of AGV$_i$ at $t=j$. An additional schedule dictionary is kept with for each job the AGV that completes it and when this happens.

\resizebox{0.98\textwidth}{!} {%
	\begin{minipage}{0.35\textwidth}
		\[
		\text{Routing} =
		\begin{bmatrix}
			v_{0,0} & v_{0,1} & v_{0,2} & \ldots \\
			v_{1,0} & v_{1,1} & v_{1,2} & \ldots \\
			\vdots & \vdots & \vdots & \ddots
		\end{bmatrix}
		\]
	\end{minipage}
	\hfill
	\begin{minipage}{0.65\textwidth}
		\[
		\text{Scheduling} = \left\{ 
		\begin{aligned}
			&\text{Job}_1 & \rightarrow & \, [\text{AGV}, t_{\text{start}}, t_{\text{end}}] \\
			&\text{Job}_2 & \rightarrow & \, [\text{AGV}, t_{\text{start}}, t_{\text{end}}] \\
			& & \vdots & \\
			&\text{Job}_n & \rightarrow & \, [\text{AGV}, t_{\text{start}}, t_{\text{end}}]
		\end{aligned}
		\right\}
		\]
\end{minipage}}

A feasible solution can be generated using one of the heuristics in Subsection~\ref{sec:heuristic}. A time limit and a maximum amount of iterations without improvement were used as stop criteria. The tabu list was implemented using a first-in-first-out queue and a set. Moves that result in a new global best cost are used as aspiration criterion, which means they can override the tabu state of that move. The following moves were incorporated:
\begin{itemize}
	\item Assigning or unassigning a job $j$. Partial (un)assignments are allowed:
	\[
	\text{Scheduling}[\text{Job}_j] = [\text{AGV}, t_{\text{start}}, t_{\text{end}}] \rightarrow [\text{AGV}, t_{\text{start}}, \text{null}]
	\]
	
	\item Changing a single node in the path of an AGV $i$ in such a way that it still results in a continuous path:
	\[
	\text{Routing}[i] = [v_0, v_0, v_1, v_2, \cdots] \rightarrow [v_0, v_1, v_1, v_2, \cdots]
	\]
	
	\item Shifting an entire loop of AGV $i$ forward or backward in time. This loop may contain some job $j$:
	\[
	\text{Routing}[i] = [v_0, v_0, v_1, v_2, v_2, v_3, v_0, v_0] \rightarrow [v_0, v_1, v_2, v_2, v_3, v_0, v_0, v_0]
	\]
	\[
	\text{Scheduling}[\text{Job}_j] = [\text{AGV}_i, 3, 6] \rightarrow [\text{AGV}_i, 2, 5]
	\]
	
	\item Unassigning or reassigning an entire loop of AGV $i$:
	\[
	\text{Routing}[i] = [v_0, v_0, v_1, v_2, v_2, v_3, v_0, v_0] \rightarrow [v_0, v_0, v_0, v_0, v_0, v_0, v_0, v_0]
	\]
\end{itemize}

Additionally, these operations were further pruned by enforcing correct loading and unloading behavior with regard to a single AGV. For example, an operation that results in an AGV moving while (un)loading gets pruned. Additionally, operations resulting in paired jobs being completed in the wrong order were also removed. This pruning was deemed necessary as TS would otherwise create a schedule with conflicts that are too difficult to resolve.

The cost function was based on violated constraints as described in Subsections \ref{sec:MathForm} and \ref{sec:mathOnline}. This cost function was further enriched with the following extra terms in order to steer the algorithm to promising regions of the solution space:
\begin{itemize}
	\item $R_1$: The amount of nodes present in the current solution where an unassigned job needs to start or end.
	\item $R_2$: The amount of extra time steps used to complete a job compared to the shortest possible.
	\item $R_3$: The amount of time steps an AGV is idle at the end of a solution.
	\item $R_4$: The amount of time steps an AGV is idle at the start of a solution.
	\item $R_5$: Whether or not both parts of a job pair were done by a single AGV.
\end{itemize}
This results in the following cost function which uses $w_i \in \mathbb{N}_0$ and $W_i \in \mathbb{Z}$ as weights:

$\texttt{cost} = \sum_i w_i \cdot \texttt{constraint}_i + \sum_i W_i \cdot R_i$.

Once a feasible solution is found, the last column of the routing matrix is removed, all jobs that ended at this final time step get partially unassigned and the search is restarted.

\section{Modeling a real plant}
\label{sec:modeling}
For the online scenario, instances were created using historical job-data and the layout of a real plant owned by a manufacturing company. This plant has seven AGVs with two slots each and they currently use a greedy strategy similar to the one described in Section~\ref{sec:Greedy} to assign jobs to the AGVs.

\begin{figure}
	\centering
	\includegraphics[width=\textwidth, ]{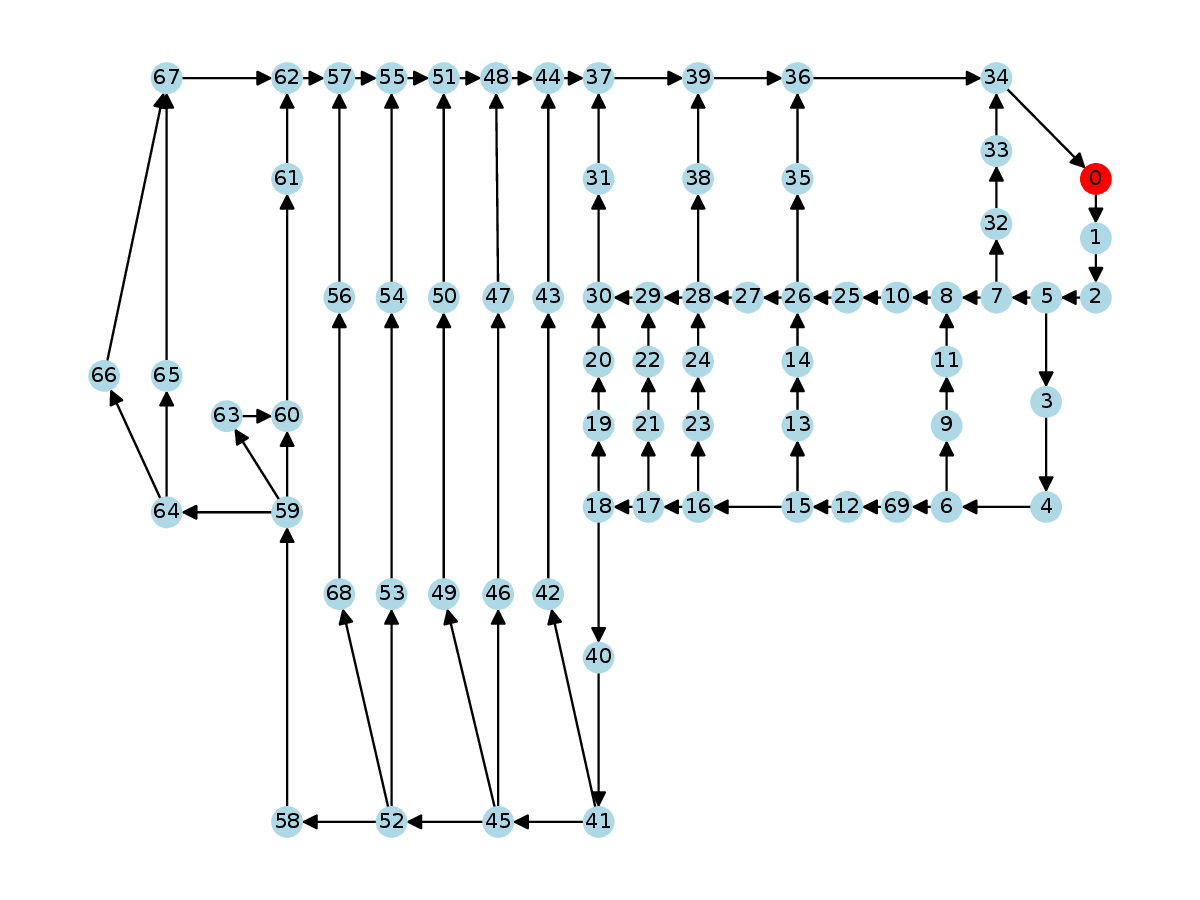}
	\caption{The simplified directed graph based on a map of a real facility and historical job-data. }
	\label{fig:prereduction}
\end{figure}

No real travel times between locations in the layout were given, so this was extrapolated based on the historical job-data. This information combined with the layout was then merged into a single graph with 320 nodes. This graph was then simplified to reduce computational complexity by merging nodes in such a way that the travel time between these newly merged nodes was approximately 20 seconds. The resulting directed graph has 70 nodes and can be seen in Fig.~\ref{fig:prereduction}. Each directed edge represents one time step of 20 seconds. Node zero is the stockroom, where AGVs can load new full pallets or unload empty ones. Directed edges connecting a node to itself are not drawn. A solution for this simplified version can be converted to one for the original graph by unmerging all these nodes. A path passing through a merged node then simply traverses these unmerged nodes in order.

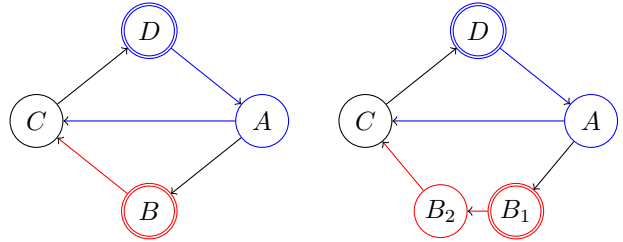
\begin{figure}
	\begin{subfigure}[t]{0.5\textwidth}
		\centering
		\begin{tikzpicture}
			\node[shape=circle,draw=blue, inner sep=0pt, minimum size=20pt] (A) at (3,0) {$A$};
			
			\node[shape=circle,draw=red, inner sep=0pt, minimum size=20pt, accepting] (B) at (1.5, -1.2) {$B$};
			\node[shape=circle,draw=black, inner sep=0pt, minimum size=20pt] (C) at (0,0) {$C$} ;
			\node[shape=circle,draw=blue, inner sep=0pt, minimum size=20pt, accepting] (D) at (1.5,1.2) {$D$} ;
			
			\path [->] (A) edge node[left] {} (B);
			\path [->] (A) edge[draw=blue] node[left] {} (C);
			\path [->] (B) edge[draw=red] node[left] {} (C);
			\path [->] (C) edge node[left] {} (D);
			\path [->] (D) edge[draw=blue] node[left] {} (A);
		\end{tikzpicture}	
		\caption{Graph before unmerge.}
	\end{subfigure}
	\begin{subfigure}[t]{0.5\textwidth}
		\centering
		\begin{tikzpicture}
			\node[shape=circle,draw=blue, inner sep=0pt, minimum size=20pt] (A) at (3,0) {$A$};
			
			\node[shape=circle,draw=red, inner sep=0pt, minimum size=20pt] (B2) at (1, -1.2) {$B_2$};
			\node[shape=circle,draw=red, inner sep=0pt, minimum size=20pt, accepting] (B1) at (2, -1.2) {$B_1$};
			\node[shape=circle,draw=black, inner sep=0pt, minimum size=20pt] (C) at (0,0) {$C$} ;
			\node[shape=circle,draw=blue, inner sep=0pt, minimum size=20pt, accepting] (D) at (1.5,1.2) {$D$} ;
			
			\path [->] (A) edge[draw] node[left] {} (B1);
			\path [->] (B1) edge[draw=red] node[left] {} (B2);
			\path [->] (A) edge[draw=blue] node[left] {} (C);
			\path [->] (B2) edge[draw=red] node[left] {} (C);
			\path [->] (C) edge node[left] {} (D);
			\path [->] (D) edge[draw=blue] node[left] {} (A);
		\end{tikzpicture}
		\caption{Graph after unmerge.}
	\end{subfigure}
	\caption{Two graphs depicting conflicts that arise when unmerging nodes.}
	\label{fig:unmerge}
\end{figure}

Note that this merging has a side effect: two unrelated jobs may share a non-stockroom start- or end-node. One solution would be to increase the capacities of these nodes and allow for parallel loading and unloading. However, this results in situations that are not physically possible in the real manufacturing plant. More specifically, AGVs are not allowed to overtake one another, which means we need to keep track of the order in which they enter and leave a merged node with a higher capacity. This rule also affects parallel loading and unloading, since AGVs would need to enter the merged node in the correct order to be able to do so.

Another solution would be to unmerge these nodes immediately, which is depicted by Fig.~\ref{fig:unmerge}. Here, an AGV is present in node $D$ and node $B$ ($B_1$). Both AGVs want to travel to node $C$ using the path marked by their respective colour. If node $B$ is unmerged into $B_1$ and $B_2$, then both AGVs suddenly arrive at the same time, which may violate the capacity of that node. This conflict is difficult to solve, but can be avoided by only unmerging nodes if no AGVs use them in their path. Since the algorithms in Section \ref{sec:Alg} only reuse the current loop of all AGVs, it is sufficient to check if this node is being used in any of these loops and not the entire solution. If it is not being used, the node can be unmerged. If it is being used, the new job that points to the node may not be added until the node can be unmerged or the other job has been completed. Note that this unmerging also makes the graph less accurate, due to an extra time step being needed for the same distance.

The dynamic unmerge was chosen since it does not directly violate physical constraints. As an added benefit, the exact method does not need any modifications since this approach modifies the underlying graph. A solution for a graph with these unmerged nodes can still be converted to the original full graph using by once again unmerging all nodes. However, AGVs with paths through nodes that were already partially unmerged will now need to wait there for some amount of time representing the additional time step introduced in the simplified model.

\section{Results} \label{sec:Results}
The algorithms described in Section~\ref{sec:Alg} were tested in an offline and online scenario. The loops heuristic was used to generate an initial solution for both the exact method and TS. Everything was implemented in Python, using the MIP-solver CBC by \cite{cbc} for the exact method. The weights for the TS cost function as described in Subsection \ref{sec:TS} were chosen using SMAC3, which was developed by \cite{JMLR:v23:21-0888}. It was given 1000 trials spread over five representative instances. The resulting weights can be seen in Table \ref{tab:weights}.

\begin{table}
	\centering
	\caption{The weights chosen by SMAC3 for TS.}
	\label{tab:weights}
	\begin{tabular}{l c c c}
		\toprule
		\multicolumn{2}{c}{Violated constraint} & \multicolumn{2}{c}{Extra terms} \\ \cmidrule(lr){1-2}  \cmidrule(lr){3-4}
		Description & Weight $w_i$ & $R_i$ & Weight $W_i$ \\
		\midrule
		Movement conflicts & 1 & $R_1$& -6\\
		Unassigned jobs & 10 & $R_2$& 1\\
		AGV capacity exceeded & 5 & $R_3$& -10\\
		Simultaneous (un)loading & 5 & $R_4$& 10\\
		& & $R_5$& 6\\ \bottomrule
	\end{tabular}
\end{table}

The quality of a solution using a set of weights was determined using the amount of time steps used. The experiments were executed on a machine with an AMD Ryzen~7 7800X3D and 32GB of RAM. The following key performance indicators were used:
\begin{itemize}
	\item Median completion time (MCT): the median of the time in minutes or time steps between a job request being added and it being completed. This only takes jobs that deliver a pallet to a non-stockroom location into account, as this is the metric of interest for the optimization of the production process.
	\item Average slot usage (ASU): indicates the efficiency of AGV's slot use, which can be calculated as the time spent holding some amount of jobs divided by the total non-idle time.
	\item Time: the amount of processing time in seconds spent to calculate the solution.
\end{itemize}

While the ASU may not be a direct optimization target, it can give an indication of how well the system can respond to newly added jobs. If AGVs utilize their slots more effectively now, more AGVs might be available for future use.

\subsection{Offline}
\label{sec:Offline}
We can use the algorithms described in Section~\ref{sec:Alg} in an offline scenario by making all requests available at $t=0$. The graph in Fig.~\ref{fig:prereduction} was used as the layout and job locations were randomly picked. These jobs are mostly paired, since this resembles the distribution in the online scenario. The specific instances can be found in Table \ref{tab:instances}. An exact solution was calculated if possible within a time frame of 20 minutes and time limit of 120 seconds was used for TS. The results of these runs can be found in Table~\ref{tab:offline}.

\begin{sidewaystable}
	
	\caption{The results of the offline experiments. Each instance has J jobs, A AGVs with a capacity of two and P percent of the jobs were paired. Every instance was run once with every algorithm. The best performing algorithm, which was picked based on MCT, then ASU and finally Time, is marked in bold for each instance.}
	\label{tab:offline}
	\resizebox{\textwidth}{!} {%
		\begin{tabular}{r r r  r r r r   r r r r  r r r r  r r r r}
			\toprule
			&   &    &      \multicolumn{4}{c}{Exact}      & \multicolumn{4}{c}{Greedy} & \multicolumn{4}{c}{TS} &      \multicolumn{4}{c}{Loops}      \\ \cmidrule
			(lr){4-7}\cmidrule(lr){8-11}\cmidrule(lr){12-15}\cmidrule(lr){16-19}
			J & A & P  & MCT  & $\sigma_{\text{CT}}$     & ASU       & Time        & MCT   & $\sigma_{\text{CT}}$     & ASU  & Time        & MCT   & $\sigma_{\text{CT}}$     & ASU  & Time    & MCT  & $\sigma_{\text{CT}}$            & ASU       & Time       \\ \midrule
			4 & 1 & 0  & \bf{22.0} &\bf{13.16}& \bf{1.14} & \bf{18.980} & 29.5  &20.99& 0.66 & 0.002       & 28.0  &14.65& 1.31 & 1.839   & 28.0       &14.65& 1.31      & 0.001      \\
			6 & 1 & 33 & 37.5      &19.74& 1.45      & 1212.771    & 48.0  &37.40& 0.76 & 0.003       & 37.5  &19.74& 1.45 & 6.265   & \bf{37.5}  &\bf{19.74}& \bf{1.45} & \bf{0.001} \\
			8 & 1 & 50 & 63.0      &38.75& 1.30       & 1212.556    & 70.0  &48.35& 0.82 & 0.002       & 63.0  &38.75& 1.30  & 22.032  & \bf{63.0}  &\bf{38.75}& \bf{1.30}  & \bf{0.001} \\
			16 & 1 & 75 & 109.5     &81.51& 1.42      & 1250.083    & 150.5 &106.33& 0.92 & 0.005       & 109.5 &81.51& 1.42 & 120.085 & \bf{109.5} &\bf{81.51}& \bf{1.42} & \bf{0.006} \\
			32 & 1 & 87 & 243.5     &180.60& 1.39      & 1233.989    & 340.0 &222.80& 0.96 & 0.047       & 243.5 &180.60& 1.39 & 123.708 & \bf{243.5} &\bf{180.60}& \bf{1.39} & \bf{0.027} \\
			48 & 1 & 75 & 389.5     &244.78& 1.42      & 1202.416    & 537.5 &311.80& 0.92 & 0.025       & 389.5 &244.78& 1.42 & 120.431 & \bf{389.5} &\bf{244.78}& \bf{1.42} & \bf{0.084} \\
			69 & 1 & 82 & 501.0     &351.81& 1.44      & 1258.328    & 791.0 &471.10& 0.94 & 0.047       & 501.0 &351.81& 1.44 & 123.548 & \bf{501.0} &\bf{351.81}& \bf{1.44} & \bf{0.226} \\ \hdashline[1pt/1.5pt]
			\phantom{\rule{1pt}{10pt}}4 & 2 & 0  & \bf{14.0} &\bf{2.87}& \bf{1.08} & \bf{11.161} & 15.0  &9.53& 0.62 & 0.002       & 15.0  &9.53& 0.62 & 1.184   & 17.5       &6.04& 1.23      & 0.001      \\
			6 & 2 & 33 & 27.0      &10.77& 1.18      & 1206.167    & 28.0  &18.70& 0.74 & 0.002       & 27.0  &10.77& 1.18 & 4.416   & \bf{27.0}       &\bf{10.77}& \bf{1.18}      & \bf{0.001}      \\
			8 & 2 & 50 & 35.0      &20.91& 1.09      & 1211.147    & 37.5  &24.19& 0.76 & 0.002       & 35.0  &20.91& 1.09 & 12.164  & \bf{35.0}       &\bf{20.91}& \bf{1.09}      & \bf{0.001}      \\
			16 & 2 & 75 & 84.5      &43.28& 1.28      & 1206.021    & 77.5  &53.72& 0.91 & 0.005       & 84.5  &43.28& 1.28 & 120.054 & \bf{84.5}       &\bf{43.28}&\bf{1.28}      & \bf{0.006}      \\
			32 & 2 & 87 & 126.0     &91.32& 1.39      & 1239.396    & 173.5 &112.25& 0.93 & 0.012       & 126.0 &91.32& 1.39 & 121.324 & \bf{126.0}     &\bf{91.32}& \bf{1.39}      & \bf{0.028}      \\
			48 & 2 & 75 & 191.0     &119.69& 1.39      & 1245.214    & 270.5 &156.55& 0.92 & 0.023       & 191.0 &119.69& 1.39 & 125.017 & \bf{191.0}      &\bf{119.69}& \bf{1.39}      & \bf{0.093}      \\
			69 & 2 & 82 & 273.0     &170.93& 1.47      & 1254.950     & 397.0 &236.73& 0.93 & 0.075       & \bf{265.0} &\bf{169.39}& \bf{1.40} &\bf{128.533} & 273.0      &170.93& 1.47      & 0.232     \\ \hdashline[1pt/1.5pt]
			\phantom{\rule{1pt}{10pt}}4 & 5 & 0  & 11.0      &2.24& 0.35      & 242.909     & \bf{11.0}  &\bf{2.24}& \bf{0.35} & \bf{0.001}       & 11.0  &2.24& 0.35 & 2.332   & 17.5       &6.04& 0.49      & 0.001      \\
			6 & 5 & 33 & \bf{13.5}      &\bf{5.24}& \bf{0.69}      & \bf{1203.120}     & 14.5  &8.17& 0.51 & 0.002       & 16.0  &3.97& 0.94 & 8.581   & 20.0       &4.75& 0.95      & 0.001      \\
			8 & 5 & 50 & 19.0      &7.36& 0.97      & 1204.617    & 20.5  &11.24& 0.55 & 0.002       & 19.0  &7.36& 0.97 & 9.716   & \bf{19.0}       &\bf{7.36}& \bf{0.97}      & \bf{0.002}      \\
			16 & 5 & 75 & 47.0      &17.55& 1.08      & 1216.964    & 57.0  &21.80& 0.83 & 0.005       & 47.0  &17.55& 1.08 & 63.639  & \bf{47.0}       &\bf{17.55}& \bf{1.08}      & \bf{0.009}      \\
			32 & 5 & 87 & 73.0      &38.32& 1.18      & 1235.578    & 92.0  &45.45& 0.86 & 0.013       & \bf{69.5}  &\bf{37.36}& \bf{1.13} & \bf{120.525} & 73.0       &38.32& 1.18      & 0.040      \\
			48 & 5 & 75 & 96.5      &48.82& 1.32      & 1224.721    & 110.5 &62.25& 0.88 & 0.023       & \bf{93.5}  &\bf{48.05}& \bf{1.30} & \bf{121.470}  & 96.5       &48.82& 1.32      & 0.094      \\
			69 & 5 & 82 & 128.0     &71.97& 1.41      & 1256.734    & 159.0 &95.43& 0.89 & 0.044       & 128.0 &71.97& 1.41 & 124.380  & \bf{128.0}      &\bf{71.97}& \bf{1.41}      & \bf{0.239}      \\ \hdashline[1pt/1.5pt]
			\phantom{\rule{1pt}{10pt}}4 & 7 & 0  & 12.0      &2.45& 0.25      & 1203.356    & \bf{11.0}  &\bf{2.24}& \bf{0.35} & \bf{0.001}       & 17.5  &5.15& 0.35 & 3.226   & 17.5       &6.04& 0.35      & 0.001      \\
			6 & 7 & 33 & \bf{11.0}      &\bf{4.97}& \bf{0.48}      & \bf{1204.584}    & 13.5  &6.65& 0.41 & 0.002       & 16.5  &4.15& 0.50 & 10.311  & 20.0       &4.75& 0.54      & 0.001      \\
			8 & 7 & 50 & \bf{15.0}      &\bf{6.97}& \bf{0.71}      & \bf{1207.379}    & 17.5  &8.04& 0.49 & 0.003       & 17.5  &8.04& 0.49 & 26.490   & 19.0       &7.36& 0.69      & 0.003      \\
			16 & 7 & 75 & 29.5      &15.11& 0.86      & 1214.614    & 35.0  &16.14& 0.68 & 0.007       & 29.5  &15.11& 0.86 & 120.772 & \bf{29.5}       &\bf{15.11}& \bf{0.86}      & \bf{0.012}      \\
			32 & 7 & 87 & 55.0      &28.10& 1.08      & 1240.237    & 67.5  &33.31& 0.81 & 0.013       & 55.0  &28.10& 1.08 & 120.470  & \bf{55.0}       &\bf{28.10}& \bf{1.08}      & \bf{0.047}      \\
			48 & 7 & 75 & 73.5      &36.61& 1.25      & 1237.213    & 88.0  &44.88& 0.84 & 0.027       & 73.5  &36.61& 1.25 & 120.163 & \bf{73.5}       &\bf{36.61}& \bf{1.25}      & \bf{0.099}      \\
			69 & 7 & 82 & 109.0     &54.89& 1.31      & 1211.055    & 129.0 &69.61& 0.85 & 0.055       & 109.0 &54.89& 1.31 & 123.295 & \bf{109.0}      &\bf{54.89}& \bf{1.31}      & \bf{0.263}      \\     \\ \bottomrule
		\end{tabular}
	}
\end{sidewaystable}

\begin{table}
	\centering
	\caption{The instances used for the offline benchmarks. Each instance has J jobs with P percent of the jobs paired. If $k \in U_i$ then instance $i$ has an unpaired job from the stockroom to node $k$.}
	\label{tab:instances}
	\resizebox{\textwidth}{!} {%
		\begin{tabular}{c c c l l}
			\toprule
			Instance $i$ & J  & P  & Unpaired $U_i$                                  & Paired $P_i$                                                                                    \\ \midrule
			$a$      & 4  & 0  & 14, 36, 37, 69                                  & $\emptyset$                                                                                     \\
			$b$      & 6  & 33 & $U_{a}$                                         & $P_{a} \cup  \{20, 28\}$                                                                        \\
			$c$      & 8  & 50 & $U_{b}$                                         & $P_{b} \cup  \{10, 47\}$                                                                        \\
			$d$      & 16 & 75 & $U_{c}$                                         & $P_{c} \cup  \{11, 12, 39, 44, 46, 51, 55, 68\}$                                                \\
			$e$      & 32 & 87 & $U_{d}$                                         & $P_{d} \cup  \{4, 17, 21, 23, 24, 29, 30, 32, 40, 50, 52, 56, 60, 62, 63, 65\}$                 \\
			$f$      & 48 & 75 & $U_{e} \cup  \{7, 15, 18, 31, 35, 43, 57, 64\}$ & $P_{e} \cup  \{5, 8, 25, 34, 38, 53, 61, 66\}$                                                  \\
			$g$      & 69 & 82 & $U_{f}$                                         & $P_{f} \cup  \{1, 2, 3, 6, 9, 13, 16, 19, 22, 26, 27, 33, 41, 42, 45, 48, 49, 54, 58, 59, 67\}$ \\ \bottomrule
		\end{tabular}
	}	\vspace{1em}
\end{table}

\begin{table}
	\centering
	\caption{The $p$-values when comparing the expected values of the set of completion times Comp or the slot usage SU of two algorithms in the offline scenario.}
	\label{tab:resOffline}
	\begin{tabular}{c c c c c}
		\toprule
		& \multicolumn{2}{c}{Loops} & \multicolumn{2}{c}{TS} \\
		\cmidrule(lr){2-3}\cmidrule(lr){4-5} & Comp     & SU             & Comp     & SU          \\ \midrule
		Greedy                & {\raisebox{0.25\height}{\small  $<$}}1e-16 & 6e-6           & {\raisebox{0.25\height}{\small  $<$}}1e-16 & 2e-5        \\
		Loops                 &          &                & 0.034     & 0.093        \\ \bottomrule
	\end{tabular}
\end{table}

We can compare the completion times of these algorithms on a per-job basis. This is not possible for the slot usage, so here the averages will be compared instead. This comparison will be done using a Wilcoxon signed-rank test, which tests whether the two populations have an equal expected value as null-hypothesis. The results can be seen in Table~\ref{tab:resOffline}. Using $\alpha = 0.05$ as the level of significance, we can conclude there is a significant difference when comparing the expected value of the completion times, albeit only barely when comparing the loops heuristic and TS. There is also a significant difference of slot usage between the greedy heuristic and any of the other algorithms, but not between the loops heuristic and TS.

On smaller instances, the exact method is able to finish its search and dominates as a result. If there are more AGVs than jobs, such as instances 4|5|0 or 4|7|0, a greedy assignment where one AGV does one job should be optimal. In these cases, the greedy heuristic outperforms the loops heuristic.

On most larger instances, the exact method is unable to make any improvements to the initial solution provided by the loops heuristic within the 20-minute time frame. Consequently, if TS or the greedy heuristic outperform the loops heuristic, they will outperform the exact method as well.

\subsection{Online}
For the online scenario, we will use the job-data and graph described in Section~\ref{sec:modeling}. A time limit of 20 seconds per scheduling period was chosen, which corresponds to the discrete travel time between nodes in Fig.~\ref{fig:prereduction}. Each instance will have seven AGVs with a slot capacity of two, since this is representative of the real manufacturing plant. 

\subsubsection{Density-based instances}
\label{sec:density}
\begin{figure}[]
	\centering
	\begin{minipage}[t]{.48\textwidth}
		\centering
		\includegraphics[width=\textwidth, trim=3cm 0 3cm 0]{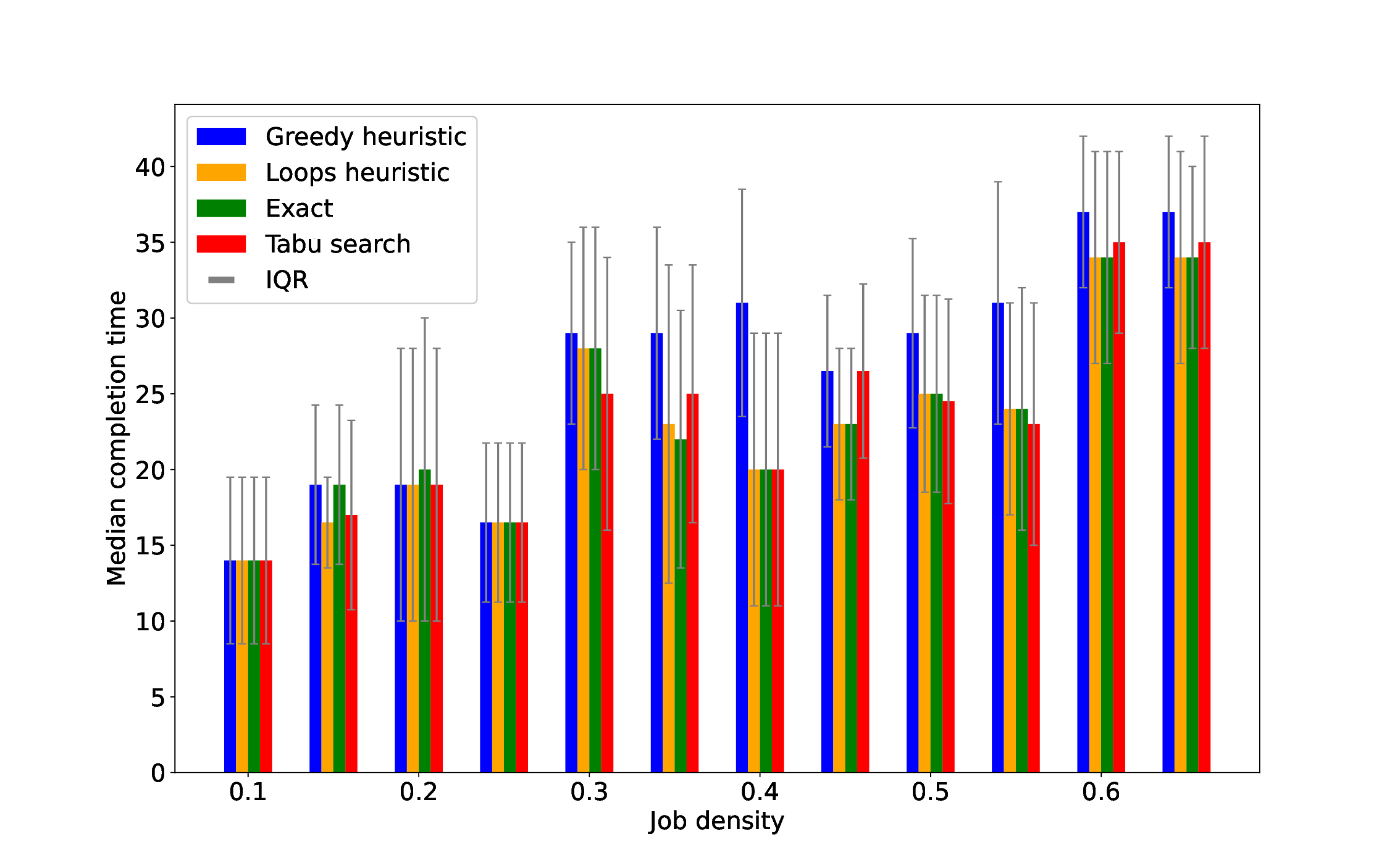}
		\caption{MCT in minutes for different instances with a given job density and a job window of 20 jobs. These instances were made using the historical data.}
		\label{fig:density_completion}
	\end{minipage}%
	\hfill
	\begin{minipage}[t]{0.48\textwidth}
		\centering
		\includegraphics[width=\textwidth, trim=3cm 0 3cm 0]{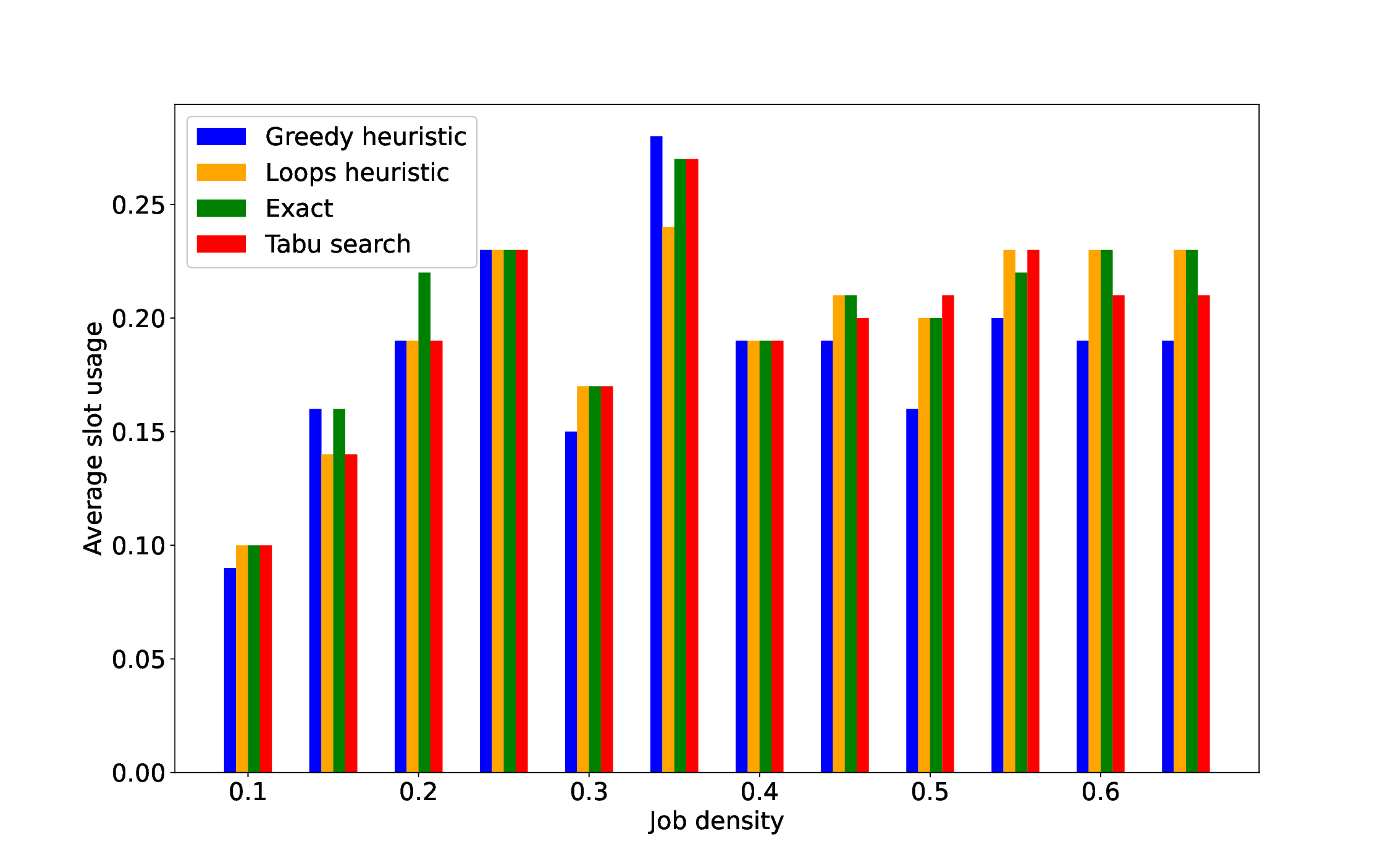}
		\caption{ASU for different instances with a given job density and a job window of 20 jobs. These instances were made using the historical data.}
		\label{fig:density_asu}
	\end{minipage}
\end{figure}

As a first experiment, instances were created using the historical data and a job window of 20 jobs. Such a window was picked based on job density: a job density of 0.5 with a job window of 20 jobs means these jobs should be added consecutively over a period of 40 time steps. The results can be found in Fig.~\ref{fig:density_completion} and Fig.~\ref{fig:density_asu} for respectively the completion time and slot usage. Note that a better MCT does not always seem to correspond to an improved SU. Additionally, the exact method actively deteriorates the result found by the loops heuristic on specific instances. This is due to it only optimizing for the current situation and not the entire time horizon.

A Wilcoxon signed-rank test will be used to compare the different algorithms. The result can be found in Table~\ref{tab:online1}. Using $\alpha = 0.05$ as the level of significance, we can conclude there is a significant difference for the MCT when comparing any two algorithms except for the loops heuristic and the exact method. On the other hand, the SU does not show a significant difference except when comparing the greedy heuristic with the exact method.

\begin{table}
	\centering
	\caption{The $p$-values when comparing the expected value of the set of completion times Comp on a per-job basis or the slot usage SU of two algorithms in the online scenario using densities.}
	\label{tab:online1}
	\begin{tabular}{c c c c c c c}
		\toprule
		&             \multicolumn{2}{c}{Loops}              & \multicolumn{2}{c}{TS} &             \multicolumn{2}{c}{Exact}              \\ \cmidrule
		(lr){2-3}\cmidrule(lr){4-5}\cmidrule(lr){6-7} & Comp                                       & SU    & Comp  & SU             & Comp                                       & SU    \\ \midrule
		Greedy                     & {\raisebox{0.25\height}{\small  $<$}}1e-16 & 0.191 & 2e-12 & 0.064          & {\raisebox{0.25\height}{\small  $<$}}1e-16 & 0.012 \\
		Loops                     &                                            &       & 3e-5  & 0.785          & 0.2                                        & 0.144 \\
		TS                       &                                            &       &       &                & 3e-6                                       & 0.107 \\ \bottomrule
	\end{tabular}
\end{table}

\subsubsection{Full day instances}
\label{sec:fullDay}
\begin{figure}
	\centering
	\includegraphics[width=0.5\linewidth, trim=0.5cm 0 0.5cm 0]{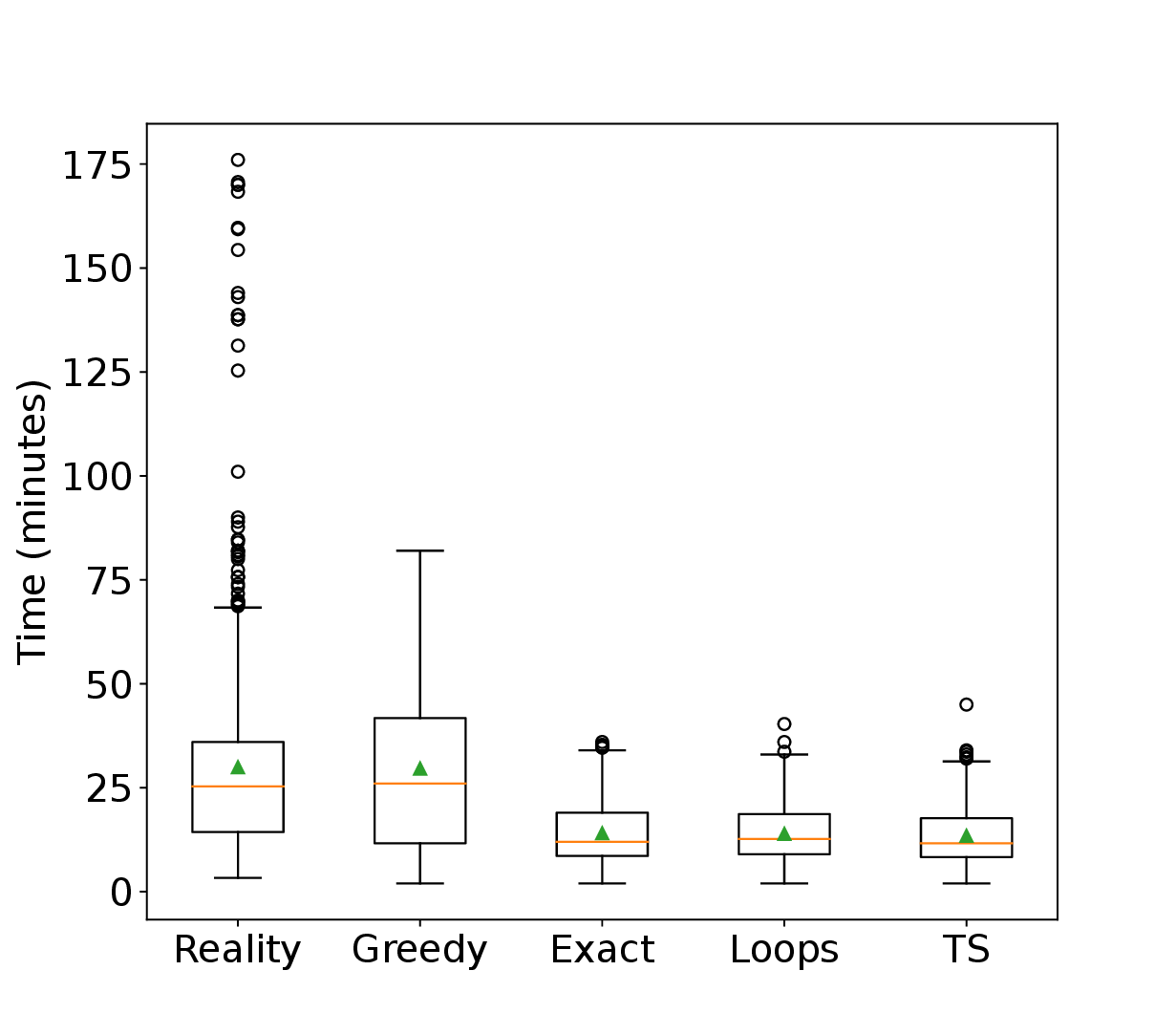}
	\caption{Box plot of the completion time in minutes for all ($n=251$) jobs for a given day in the online scenario. In the order shown above, the MCTs are 25.3, 26.0, 12.0, 12.6 and 11.7 minutes.}
	\label{fig:box}
\end{figure}

As a second experiment, the algorithms were run on an instance representing a full day worth of requests with no modifications. The completion time for each of the algorithms is given in Fig. \ref{fig:box}.

\begin{table}
	\centering
	\caption{The $p$-values when comparing the expected values of the set of completion times on a per-job basis of two algorithms in the online scenario using full day instances.}
	\label{tab:online2}
	\begin{tabular}{c c c c c}
		\toprule
		& Greedy & Exact                                      & Loops                                      & TS                                         \\ \midrule
		Reality & 0.438  & {\raisebox{0.25\height}{\small  $<$}}1e-16 & {\raisebox{0.25\height}{\small  $<$}}1e-16 & {\raisebox{0.25\height}{\small  $<$}}1e-16 \\
		Greedy  &        & {\raisebox{0.25\height}{\small  $<$}}1e-16 & {\raisebox{0.25\height}{\small  $<$}}1e-16 & {\raisebox{0.25\height}{\small  $<$}}1e-16 \\
		Exact  &        &                                            & 0.721                                      & 0.181                                      \\
		Loops  &        &                                            &                                            & 0.067                                      \\ \bottomrule
	\end{tabular}
\end{table}

Statistical tests on the slot usage are not useful here, since this is just a single value for each instance. However, we can once again test the expected value for the completion time on a per-job basis. The appropriate test in this case is once again a Wilcoxon signed-rank test with the null-hypothesis that the expected values are equal. The resulting $p$-values can be found in Table~\ref{tab:online2}. The ASU could not be extrapolated from the historical data, but was 0.49, 0.49, 0.48 and 0.50 for respectively the greedy heuristic, the exact method, the loops heuristic and TS.

We can conclude, using $\alpha = 0.05$ as the level of significance, that our model and greedy implementation are representative for the actual manufacturing plant. Furthermore, there is a statistically significant difference in the MCT between these two and the exact method, loops heuristic or TS. More concretely, the average job is completed more than twice as fast when using any of these algorithms instead of the greedy heuristic (26.0 vs 12.0, 12.6 or 11.7 minutes respectively). While there is a difference between the MCTs of the loops heuristic and TS, this difference is barely statistically insignificant with a $p$-value of 0.067.

Note that all three non-greedy algorithms produce significantly less outliers, which results in less jobs being greatly overdue. This is reflected in the standard deviation of the completion time, with values of 74.06, 43.95, 18.29, 17.43 and 16.95 for respectively the real data, greedy heuristic, exact method, loops heuristic and TS.

\section{Threats to validity}\label{sec:Threats to validity}
While our proposed solution shows promising results in the context of an idealized production floor environment, there are several external factors that can affect its performance when implemented in a real-world setting. One such example would be an AGV breaking down or malfunctioning. Even something as simple as a person blocking the path of an AGV for some amount of time can render the scheduling and routing plan invalid. We can mitigate this threat by implementing the algorithm used in the actual manufacturing plant. A statistical comparison between the historical job-data and the algorithms' output using the model can then be done, which can be found in Subsection~\ref{sec:fullDay}. While there was no statistically significant difference for this manufacturing plant, this may not always hold true.

A second threat to validity is the way we compare these algorithms using the KPIs. From a pure optimization perspective, we are only interested in minimizing the MCT. However, from an operator's perspective, a high ASU as secondary target implies more AGVs might be available for future jobs. The exact method directly optimizes for the MCT, while the loops heuristic and TS both use multiple different indirect metrics. One could argue that the results are therefore incomparable, since they all use different objectives. It would even be possible to artificially inflate the ASU by holding on to empty pallets as these are excluded from the MCT. Nonetheless, while TS and the loops heuristic might not directly optimize for the MCT, the way they are implemented does. For example, the weights for TS were chosen based on MCT regardless of ASU. The loops heuristic uses ASU as the last criterion when all others are equal and tries to load and unload as early as possible. As a result, both of them should prioritize the MCT and should not be able to unfairly inflate their ASU.

Another potential threat to validity is the limited processing time of the algorithms due to the online nature of the problem. Both the exact method and TS were heavily constrained due to this, i.e. both of them routinely needed more than 20 seconds in the offline scenario which can be seen in Table~\ref{tab:offline}. This results in them underperforming in the online scenario, where the per-period time limit was 20 seconds. This means that a more efficient implementation in a potentially faster language or even simply better hardware could yield different results.

A final threat to validity is proving the solutions found by the algorithms do not violate any constraints. Since we already have a mathematical formulation and corresponding implementation of these constraints, we can simply convert a found solution to an assignment of the decision variables as described in Section~\ref{sec:MathForm} or \ref{sec:mathOnline} and verify the result with a MIP solver, such as the one by \cite{cbc}. However, this strategy does not work in the online case, due to the unmerging of nodes. Instead of checking the entire solution at the end, we can verify the partial result of every scheduling period separately.

\section{Conclusion and future works}\label{sec:conclusion}
In this paper, we proposed a novel loops heuristic to solve the online scheduling and routing problem for automated guided vehicles (AGVs). We compared this new method to adapted versions of already existing solving strategies in both offline and online scenarios. The online scenario made use of real-life data, which had to be modelled. We provided a way to somewhat simplify this model.

The loops heuristic almost always outperformed a greedy implementation. In the offline scenario, it performed worse than TS, albeit only barely. In the online scenario, it outperformed TS in density based instances and found a similar solution in full day instances. Note that the new method is orders of magnitude faster than TS in all cases.

Future works might try to incorporate a predictive component that makes AGVs wait if a combinable job request is predicted to be added soon. The new method could also make use of variable branching, which can depend on how much time is available. The modelling method could also be improved by adding a way to re-merge nodes when possible.

\bibliographystyle{abbrvnat}
\bibliography{bib}

\section*{Statements \& Declarations}
The authors would like to thank Jorik Jooken for helpfull insights and motivating discussions about the subject. 

\subsection*{Funding}
The research of Jan Goedgebeur was supported by internal funds of KU Leuven. 

\subsection*{Competing interests}
The authors have no relevant financial or non-financial interests to disclose.

\subsection*{Data availability}
The dataset used for the offline instances are available in Table \ref{tab:instances}. The dataset used for the online instances are not publicly available due to a non-disclosure agreement with the manufacturing company that provided their historical job-data.

\end{document}